# Super interference fringes of two-photon photoluminescence in individual Au nanoparticles: the critical role of the intermediate state

Yao Li,[1,2,†] Yonggang Yang,[1,2,†] Chengbing Qin,[1,2,*] Yunrui Song,[1,2] Shuangping Han,[1,2] Guofeng Zhang,[1,2] Ruiyun Chen,[1,2] Jianyong Hu,[1,2] Liantuan Xiao,[1,2,*] and Suotang Jia[1,2]

[1] State Key Laboratory of Quantum Optics and Quantum Optics Devices, Institute of Laser Spectroscopy, Shanxi University, Taiyuan, Shanxi 030006, China.

[2.] Collaborative Innovation Center of Extreme Optics, Shanxi University, Taiyuan, Shanxi 030006, China.

[†] These authors contributed equally to this work.
[*] Author to whom correspondence should be addressed.

Chengbing Qin, chbqin@sxu.edu.cn

Liantuan Xiao, xlt@sxu.edu.cn

**Abstract:**

The interaction between light and metal nanoparticles enables investigations of microscopic phenomena on nanometer length and ultrashort time scales, benefiting from strong confinement and enhancement of the optical field. However, the ultrafast dynamics of these nanoparticles are primarily investigated by multiphoton photoluminescence on picoseconds or photoemission on femtoseconds independently. Here, we presented two-photon photoluminescence (TPPL) measurements on individual Au nanobipyramids (AuNP) to reveal their ultrafast dynamics by two-pulse excitation on a global time scale ranging from sub-femtosecond to tens of picoseconds. Two-orders-of-magnitude photoluminescence enhancement, namely super interference fringes, has been demonstrated on tens of femtoseconds. Power-dependent measurements uncovered the transform of the nonlinearity from 1 to 2 when the interpulse delay varied from tens of femtoseconds to tens of picoseconds. We proved that the real intermediate state plays a critical role in the observed phenomena, supported by numerical simulations with a three eigenstates model and further experiments on Au nanospheres with different diameters. The crucial parameters, including the dephasing time, the radiative rate, and the coupling between different states, have been estimated using numerical simulations. Our results provide insight into the role of intermediate states in the ultrafast dynamics of noble metal nanoparticles. The giant photoluminescence in super interference fringes enables potential practical applications in imaging, sensing, and nanophotonics.

## 1. Introduction

Noble metal nanostructures, such as Au and Ag nanoparticles, have attracted much research interest in the past decades due to their localized surface plasmon resonance (LSPR), representing coherent collective oscillations of the conduction electrons at the surface of metallic nanostructures[1,2]. This collective oscillation presents two intimated features: sub-wavelength confinement of optical field and thus giant enhancement of localized electric field[3,4], both of which are important for many promising applications, such as sensing[5,6], lasing[7,8], energy harvesting[9,10], and photothermal cancer therapy[11,12]. For example, near-field confinement of the local field has been used to break the light diffraction. Up to now, the spatial resolution below one nanometer has been achieved by matching the resonance of the nanocavity plasmon to the molecular vibronic transitions[13,14]. On the other hand, the dramatically enhanced light field has been utilized to prompt the performance of nonlinear optical effects, optical trapping, and surface-enhanced Raman scattering (SERS)[15-18]. The magnitude of field enhancement by LSPR is sensitive to various parameters, such as the morphology of metal nanoparticles, the surrounding dielectric media, and the interparticle plasmon coupling[19,20]. A significant research effort currently focused on the coherent control of femtosecond energy localization in nanosystems by selectively exciting a number of eigenmodes of the metal nanoparticles through adaptive shaping of the laser pulse phase and amplitude [21]. However, full control of nano-optical fields is still challenging due to the complex dynamics of the system, which involves broad time scales from several femtoseconds to picoseconds[22]. Thus, exploring their ultrafast dynamics and understanding the relevant mechanisms are crucial for further applications.

Efforts toward the time-resolved measurements of metal-nanoparticles dynamics have involved two approaches. One investigated the dephasing processes on the time scale of several to tens of femtoseconds through traditionally nonlinear interferometric autocorrelation (IAC) measurements. Experiments based on high-order harmonic generation (such as second harmonic generation, SHG, and third harmonic generation, THG) and multiphoton photoemission (MPPE) processes have been carried out[23-27]. According to the reconstructed plasmon-enhanced electric field $E(t)$ from these measurements, the pulse duration and thus the dephasing time can be derived. On the other aspect, the hot-carrier relaxation dynamics on the time scales of picoseconds have been studied by multiphoton photoluminescence (MPPL) [28-30]. The temporal response of the carriers' relaxation extending from sub- to several picoseconds has been reported. Notwithstanding, the dynamics of metal nanoparticles have been severally performed through two individual approaches, and the corresponding ultrafast behaviors have been discussed independently. However, the global investigation, involving time scales from sub-femtosecond to tens of picoseconds through the same approach, is highly desired to evaluate the unique phenomena and uncover the underlying mechanisms.

In this work, we address this limitation through two-photon photoluminescence (TPPL) measurements with a homemade Michelson interferometer by using ultrashort femtosecond laser pulses. Ultrafast dynamics of individual Au nanobipyramids (AuNP) on a global time scale—from sub-femtosecond to tens of picoseconds—have been measured and analyzed. Almost two orders of magnitude photoluminescence enhancement, comparied to traditional TPPL, have been determined on tens of femtoseconds' time scale. We named this phenomenon as super interference fringes. The decay of TPPL, from tens of femtoseconds to tens of picoseconds, was similar to the previous works. The nonlinearity orders of 2 have been determined for the global time scales when varying the total laser power. Interestingly, linear power-dependent behavior at zero interpulse delays was first obtained when changing the incident power of the first or second laser pulse. We have proved that the real intermediate state is critical for the exploring of super interference fringes and the linear optical behavior, which have been further supported by numerical simulations and extended experiments on other Au nanospheres with different diameters. The dephasing rate, relaxation of hot carriers, and the coherent coupling between different eigenstates are also evaluated from numerical simulations.

## 2. Results and discussion

Ultrafast dynamics of metal nanoparticles on global time scales were constructed using an inverted microscope (sketched in Figure 1a) with a broadband Ti:sapphire oscillator (Vitara, Coherent, Inc.) delivering 15 fs laser pulses centered about 800 nm at the repetition rate of 80 MHz. Spectral characteristics of the laser pulse have been presented in Figure 1b (red shaded area). Two pulse replicas were produced by a homebuilt Michelson interferometer with a stepping motor enabling interpulse delays (the time delay $\Delta t$ between the two pulses) between -15 ps to 15 ps in steps of 0.1 fs.

Hence, the time-resolved MPPL experiments with sufficiently short laser pulses and high time-interval resolution, as used in our experiment, can provide direct measurements of the dephasing process of the plasmon states. On the other aspect, sufficient long interpulse delays enable the study of relatively slower dynamics, such as hot carriers' relaxation. As a model system, individual rice-shaped AuNP (shown in the inset of Figure 1a) with a diameter of 20 nm and length of 50 nm were used to study the global ultrafast dynamics, benefiting from their high local field enhancement. AuNP, obtained by colloidal synthesis, ensures reproducible investigations, given that they are almost identical from one synthesis to another independently of the equipment and the personnel[31]. The typical extinction spectra of AuNP (gray dashed line), with a longitudinal plasmon resonance wavelength of about 700 nm, are presented in Figure 1b. More morphological and spectral characteristics of metal nanoparticles can be found in the electronic supplementary information (ESI, Figure S1, and S2). Nanoparticles are dispersed at a low density on the glass coverslip. In this fashion, the laser pulses can focus on an individual nanoparticle each time (the individual feature can be proved by the scanning electron imaging characterization after ultrafast investigations) through a high numerical aperture objective (100×, NA 1.49) with the laser spot close to the diffraction. To block the laser pulses, we employed a 700 nm short-pass filter (Chroma, ET700SP), the transparency spectrum of which is marked as the solid line in Figure 1b. After excited by the ultrashort laser pulses, the distribution of MPPL is mostly in the region of 500-700 nm (as the color shaded area shown in Figure 1b). Straightforwardly, the spectral feature of MPPL is similar to that of one-photon excitation by 532 nm continuum-wave (CW) laser (Figure S2), manifesting that at least two photons are needed to overcome the optical bandgap of AuNP to access MPPL by the femtosecond laser with the wavelength of 800 nm.

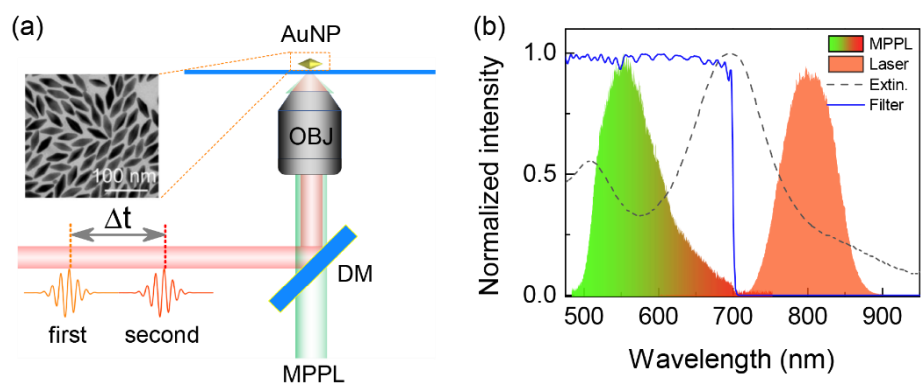

Figure 1. (a) Schematic for studying the global ultrafast dynamics of Au nanoparticles by two-pulse excitations. The inset is the transmission electron microscopy (TEM) of Au nanobipyramids (AuNP). DM, dichroic mirror; OBJ, objective. MPPL, multiphoton photoluminescence. (b) Spectral characteristics of the laser pulse (red shadow), extinction (gray dashed line), and MPPL (color shadow) of AuNP, as well as the transparent spectrum of short-pass filter (azure solid line). The intensities have been normalized by their maximum values as a guide for the eye.

The typical MPPL trace of an individual AuNP as a function of interpulse delay has been illustrated in Figure 2a. At first glance, we can find the sharp and dramatic enhancement on MPPL when the two pulses are overlapping ($\Delta t \sim 0$), comparing with MPPL obtained by either single-pulse excitation or two-pulse excitation with sufficient long interpulse delay (for example, $\Delta t \sim 10$ ps, marked as delay C). Here, the power of the single-pulse excitation was the same as that of the first or second pulse. Moreover, the intensity of MPPL gradually degenerates with the increase of the interpulse delay. To get insight into the detailed information, we plot the trace in logarithmic scale and normalize the MPPL intensity to two-pulse excitation with sufficient long interpulse (if not otherwise specified, the interpulse delay of 10 ps was used as the sufficient long interpulse delay in this context), as presented in Figure 2b. Consequently, the two-pulse MPPL is twice that of the single-pulse excitation at delay C. This result implies the coupling between the two pulses and the plasmon states of AuNP entirely vanishes. With the decrease of the interpulse delay, MPPL gradually increases, as delay B with interpulse of 2.5 ps. The change of MPPL in this process can be well fitted with a monoexponential relationship. This phenomenon was previously attributed to the residual hot electrons created by the first pulse and subsequently excited by the second pulse to produce additional MPPL, resulting in the enhancement of photoluminescence (PL) intensity[28,31]. Here, we attributed this timescale (~1.46 ps) to the lifetime of intermediate states, $\tau_1$, which will be discussed later. When the interpulse delay is close to zero, almost two orders of magnitude enhancement of PL intensity, as compared to that at delay C, can be definitely determined. Simultaneously, the minimum PL intensity is close to zero (to background). Zooming in this area (the dashed rectangle area), we can observe clear interference fringes, as shown in Figure 2c. To quantitatively describe the enhanced interference fringes, we define an enhancement factor, that is, the ratio between the maximum PL intensity and that with sufficient long interpulse delays. For this AuNP, enhancement factor is up to 102. We address this giant enhanced MPPL as super interference fringes. To our knowledge, the super interference fringes have never been discussed conscientiously in the previous works. This PL enhancement can be well reproduced on many AuNP in our experiment, with an averaged value to be 90±14 (see Figure S3).

To date, the exact mechanisms of MPPL from Au nanoparticles are still in debated, either arising from the absorption of multiple photons through virtual or real intermediate states[28,31-33], or stemming from the emission process through intraband relaxation[30,31,34]. To explore the possible mechanism, we firstly assume that AuNP absorbs photons through virtual states. Thus MPPL can be treated as a quasi-instantaneous process. The signal of IAC (*i.e.*, the MPPL intensity) can be given by $I^{AC}(\Delta t) = \int_{-\infty}^{+\infty} \left| \left( E(t) + E(t-\Delta t) \right)^n \right|^2 dt$, with $n$ being the order of nonlinearity [35,36]. Considering PL excited by CW and femtosecond lasers discussed above, two photons are enough to access MPPL. Therefore we immediately simulate the interference fingers by assuming $n$=2 according to the up-mentioned formula, as shown in Figure 2d. However, three intuitive facts illustrate the invalid of this simulation. Firstly, the enhancement factor is 8, consistent with the feature of SHG[25,37-39] rather than close to a hundred. Secondly, simulated TPPL persists for interpulse delays of less than 40 fs, rather than 60 fs in our experiment. Finally, PL decay at a longer time scale is absent (the detailed comparison can be found in Figure S4). On the other aspect, to verify the nonlinear optical behaviors stemming from intraband relaxation, we also simulate higher-order nonlinearities to match the experimental result, as shown in Figure 2e. Outwardly, the simulation seems perfectly for constructive fringes with $n$=3.8. However, PL decay is still absent. Furthermore, to illustrate the persistence of MPPL, we also display the IAC of the laser field, as shown in Figure 2f. The difference between MPPL and laser field further hints the presence of plasmon states and their relatively long dephasing time.

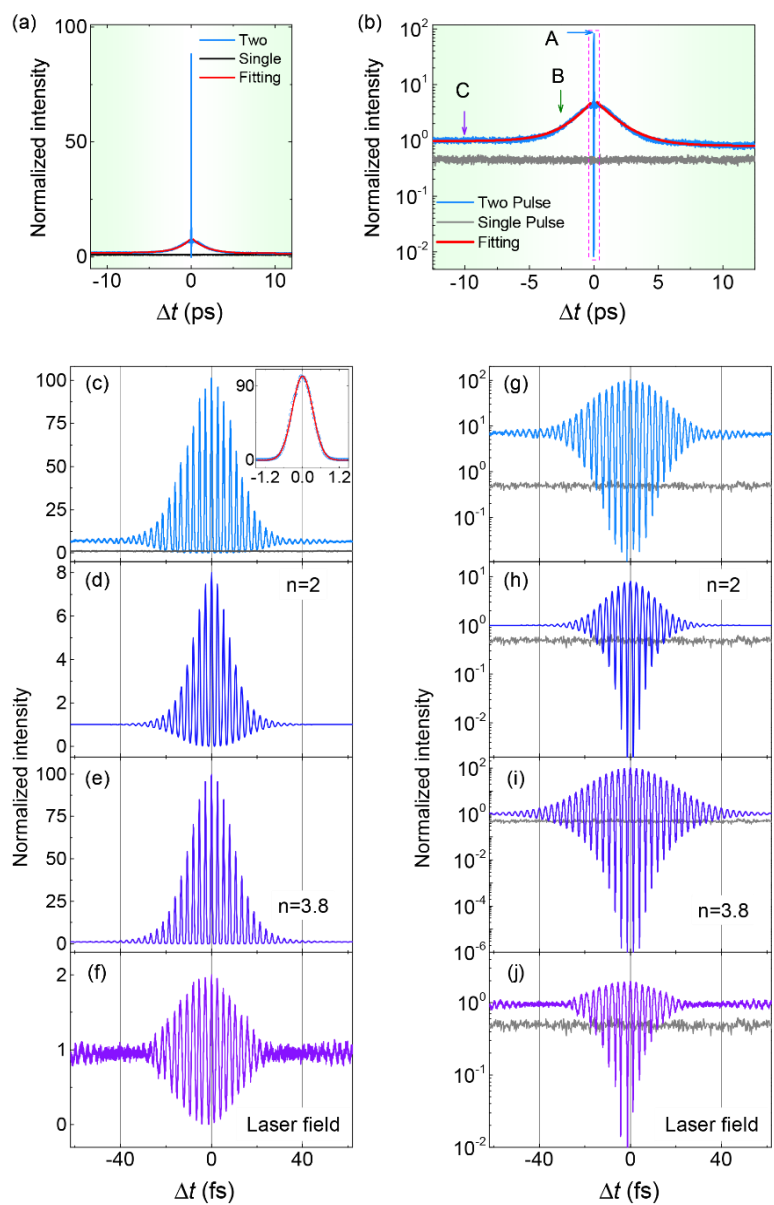

Figure 2. Ultrafast dynamics of AuNP on global time scales. MPPL of AuNP as a function of interpulse delay ($\Delta t$) in the linear (a) and logarithmic plot (b). PL intensities are normalized by the intensity of two-pulse excitation with an interpulse delay of 10 ps. The solid red lines are the monoexponential fits with a lifetime of 1.46 ps. The interference fringes of experimental MPPL, simulated MPPL with $n=2$ or $n=3.8$, and the incident laser field are presented in (c)-(f). The corresponding date in the logarithmic scales are shown in (g)-(i). The inset in c displays the profile of the interference at $\Delta t=0$, which can be fitted by Gauss function with the full width at half maximum of 0.77 fs. The trace of single-pulse MPPL is also presented as a guide for the eye.

To undoubtedly determine the nonlinearity and the underlying mechanism of MPPL, we next performed the power-dependent PL measurements at global time scales. During these measurements, the incident laser power of the first, second pulse, or the sum of them was set in the range of 5 µW to 50 µW. As we have seen, this power is one or two orders of magnitude lower than the reported works[28,31], implying the extremely high local field enhancement in our AuNP. Further increasing the laser power will result in severe random variation and then irreversible reduction in PL intensity, probably due to thermal damage to Au nanoparticles. On the global time scales investigated, we demonstrated $n=2$ by synchronously varying the first and second laser power (Figure S5), undoubtedly illustrating that the observed MPPL belongs to a two-photon process. However, as shown in Figure 3a, a novel result occurs when we explored the nonlinearity of MPPL by fixing the power of the first (or second) pulse and varying the other one. We have proved that the nonlinearity is identical for both conditions (see Figure S6). The nonlinearity is 2.06 for single-pulse excitation, coinciding with the second-order nonlinearity of TPPL. The nonlinear order at delay C is 2.04, equal to single-pulse excitation, suggesting that the coupling between the two laser pulses and the plasmon states of AuNP entirely vanishes. Interestingly, when the interpulse is close to zero, the power-dependent TPPL approaches a linear behavior, with the nonlinearity to be 0.97. When the interpulse are between delay A and C, the linear action and the two-photon process work together, presenting a regular change nonlinearity from 1 to 2, as illustrated in Figure 3b. More or less, these results agree with previous works on Au nanoantennas reported by Régis et al.[31] However, some distinct features still declare the existence of hidden dynamics. (1) The nonlinearity of PL intensity is identical, either varying the power of the first pulse or changing that of the second pulse[31]. (2) Next, the nonlinearity is strictly limited between 1 and 2. Higher nonlinearity orders are absent in our experiment[28,31]. (3) At last, the nonlinearity as a function of interpulse delay follows a monoexponential behavior (solid line shown in Figure 3b), rather than the linear relationship published[31].

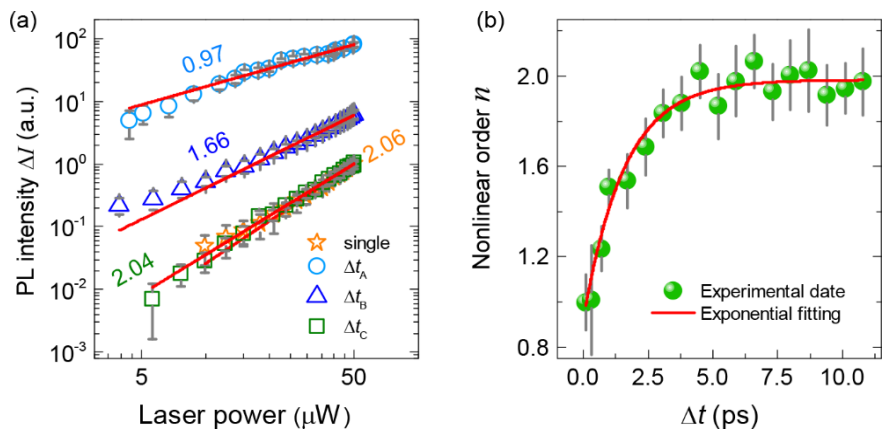

Figure 3. Power-dependent TPPL behaviors on the global time scales. (a) Power dependence PL intensity for the single-pulse excitation and the two-pulse excitations (only changing the power of the second pulse) at three interpulse delays (Delay A, B,

and C have been marked in Figure 2b). For two-pulse excitation, $\Delta I$ was determined by subtracting the background (excitation only by the first pulse) from the total intensity. The solid lines are the power-law fits, $\Delta I = a \times P^n$, with the exponents, $n$, presented in the figure. (b) Power-law exponents for two-pulse excitation as a function of interpulse delay. The solid line represents the monoexponential fits.

To interpret the super interference fringes as well as the power-dependent TPPL at the global time scale, we present a physical scenic that TPPL in AuNP arises from both two-photon absorption (TPA) through the single-pulse excitation and resonant coherent excitation through the two-pulse process involving the real intermediate states, as depicted in Figure 4a. Here we denote the ground, intermediate and excited states as |0⟩, |1⟩, and |2⟩, respectively. After the first-pulse excitation, a few electrons will be pumped to excited states (|2⟩) through TPA, resulting in relatively weak TPPL. On the other aspect, vast electrons will be pumped to the intermediate state through one-photon absorption due to their much larger absorption cross-section. These electrons will be further excited to |2⟩ to produce much stronger TPPL. However, they will relax to the ground state after excitation through the radiative and/or non-radiative process (the radiative signal is beyond the detection of our schematic). Thus, the lifetime of the intermediate state ($\tau_1$) can be determined by analyzing the TPPL decay as a function of interpulse delays, as the solid red lines shown in Figure 2b. We gain $\tau_1$ to be 1.46 ps, close to the hot-carrier relaxation through carrier-photon interactions in the previous studies of other gold nanomaterials[31,32]. Specifically, these electrons will eventually lose the phase information because of the dephasing processes such as electron-electron scattering and radiative damping. Within the dephasing time, these electrons can be coherently excited to the state $|2\rangle$ or recovered to the state $|0\rangle$, determined by the phase difference between the plasmon states and the second pulse. When the excited states are in phase with the second pulse, electrons arising from state $|1\rangle$ will be coherently excited to state $|2\rangle$ through farther one-photon absorption and thus predominate TPPL. In this case, TPPL is dramatically enhanced than pure TPA (*i.e.*, the single-pulse excitation). The extra PL intensity results in the giant enhancement on interference fingers (Figure 2c), compared to that obtained through virtual states (Figure 2d). Consequently, the power of the second laser governs the intensity of TPPL, therefore displaying a linear power-dependent behavior. Although the coherent excitation disappears beyond the dephasing time, the linear absorption of $|1\rangle \rightarrow |2\rangle$ still occupies the intensity of TPPL, and thus TPPL still emerges linear behavior on short interpulse delay ($\Delta t$=100 fs shown in Figure S7). With the increasing of interpulse delays, the ratio of linear absorption decreases, and that of TPA increases, leading to an increase of nonlinear order, as shown in Figure 3b.

To prove our model and extract the ultrafast parameters of the three eigenstates as well as their coupling, we perform a numerical simulation to match the super interference fringes. The Hamiltonian of the AuNP can be written as $H_{NP} = \sum_e E_e |e\rangle\langle e| + \sum_v E_v |v\rangle\langle v| + \sum_{ee'}\sum_{vv'} g_{ee'vv'} |e\rangle|v\rangle\langle v'|\langle e'|$, where $E_e$ and $|e\rangle$ are the eigen-energies and eigen-states of the electrons for the AuNP structure fixed at the equilibrium. $E_v$ and $|v\rangle$ are the vibrational energies and states for the nuclear vibrations at the lowest electronic state. The states $|e\rangle|v\rangle$ form a complete basis set for the Hamiltonian. As discussed above, the interaction between AuNP and the incident laser pulse $\vec{F}(t)$ can be characterized as $E_v = \vec{\mu} \cdot \vec{F}(t)$. For the two-pulse excitation, $\vec{F}(t)$ has the form of $\vec{F}(t) = \vec{F}_1(t) + \vec{F}_2(t + \Delta t)$, where the electric field of laser pulse is assumed as $\vec{F}_k(t) = \vec{F}_0 \cdot e^{-t^2/\tau_0^2} \cdot \sin(\omega_0 t)$ $(k=1,2)$. Both the first and second pulses have Gauss envelope, with the pulse duration $\tau_0$ and central frequency $\omega_0$, respectively (in the experiment $\tau_0$=15 fs; $c/(2\pi\omega_0)$=800 nm).

Considering that the strength of the laser pulse is generally very weak, we only take account of resonant processes. For this experiment, the involved states are only $E_0$, $E_1 = E_0 + \hbar\omega_{01}$, and $E_2 = E_0 + \hbar\omega_{01} + \hbar\omega_{12}$, while all the other states together with the environment can be treated either as bath degrees of freedom or as irrelevant states. The total Hamiltonian is thus transformed as $H = H_{NP} + H_B + H_{NP-B}$. The Hamiltonian of the interaction between AuNP and the laser pulse, $H_{NP}$, can be expressed as $H_{NP} = \sum_{n=0}^{2} E_n |n\rangle\langle n| - \sum_{nn'} \vec{\mu}_{nn'} \cdot \vec{F}(t)|n\rangle\langle n'|$. While the Hamiltonian of the bath ($H_B$) and its interaction with AuNP ($H_{NP-B}$) are given by $H_B = \sum_b E_b |b\rangle\langle b|$, $H_{S-B} = \sum_{nn'}\sum_{bb'}|n\rangle|b\rangle\langle b'|\langle n'|$. The other relevant states are described as bath degrees of freedom with energies $E_b$ and states $|b\rangle$. Then, the quantum state of the system can be characterized by the density operator $\rho(t) = \sum_{nn'} \rho_{nn'}(t)|n\rangle\langle n'|$. The propagation of the density operator obeys the quantum master equation $\frac{\partial}{\partial t}\rho(t) = \frac{1}{i\hbar}[H_{NP}, \rho(t)] - \Re\rho(t)$. Here $\Re$ is the dissipation super-operator defined in terms of several dephasing or relaxation rates $\Gamma_{nn'}$ as $\Re_{nn',mm'} = \delta_{nm}\delta_{n'm'}\Gamma_{nn'}$. The physical meaning of each $\Gamma_{nn'}$ is transparent. Specifically, $\Gamma_{22}$ is the sum of the bath-induced population transfer rate from $|2\rangle \to |1\rangle$ ($\gamma_{21}$) and the one from $|2\rangle \to |0\rangle$ ($\gamma_{20}$); while $\Gamma_{11}$ means the difference between the population transfer rate from $|2\rangle \to |1\rangle$ ($\gamma_{21}$) and the one from $|1\rangle \to |0\rangle$ ($\gamma_{10}$). The off-diagonal $\Gamma_{nn'}$ means bath-induced dephasing rate of the coherence between states $|n\rangle$ and $|n'\rangle$. In the numerical simulations, we define the initial condition of the quantum state as $\Gamma_{nn'}(t=-\infty) = \delta_{n0}\delta_{n'0}$.

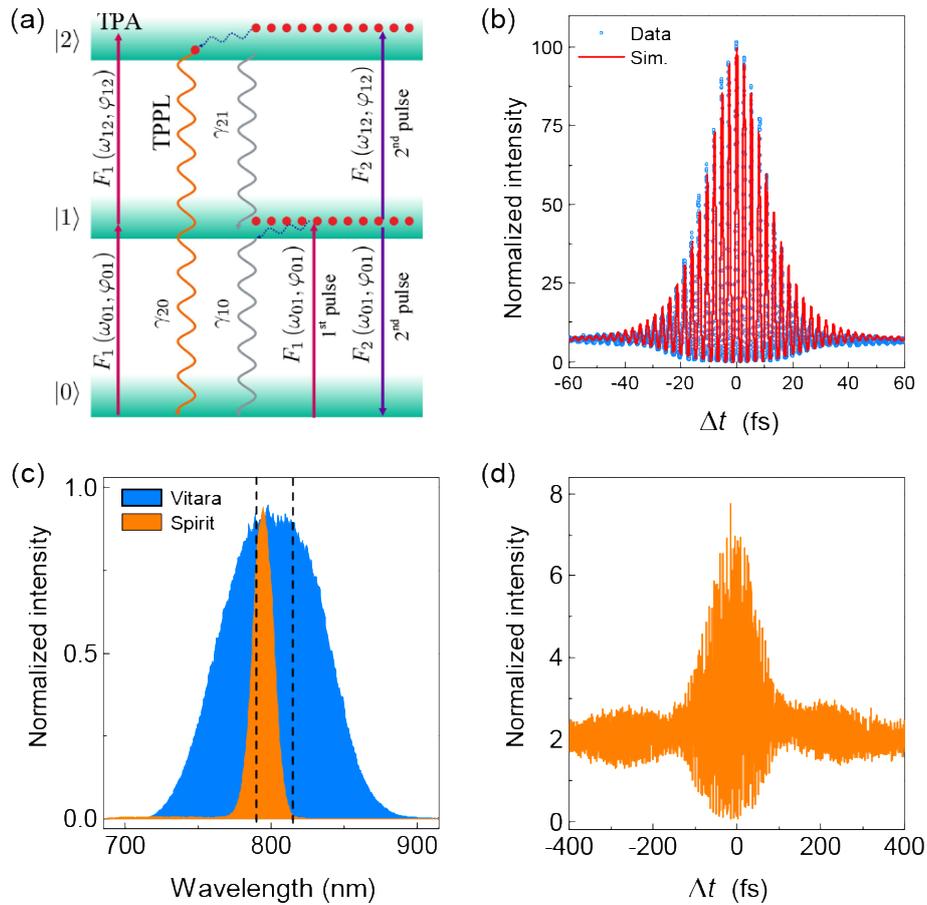

Figure 4. Sketching energy levels of AuNP and numerical simulations. (a) Schematic of the energy levels of AuNP with real intermediate states. The resonant electric field of the first and second pulses denote as $F_1$ and $F_2$, respectively. The solid and dashed wavy lines represent the radiative and relaxation processes. The red balls represent the excited electrons. TPA, two-photon absorption. (b) The experimental interference

fringes and the corresponding numerical simulation. The parameters used for simulations can be found in Table I. (c) Spectral characteristics of the broadband (Vitara) and narrowband (Spirit) femtosecond laser. Two dashed lines highlight the resonant wavelengths ($c/(2\pi\omega_1)$= 815 nm, $c/(2\pi\omega_2)$= 792 nm) for the simulation of interference in a. (d) Two-pulse excitation of the same AuNP using the narrowband laser (Spirit).

TPPL intensity is proportional to the population of the state $|2\rangle$, namely $\rho_{22}(t)$, after the laser pulse. In the numerical simulations, we use a sufficiently long time $T_f$, such that the experimental intensity is proportional to $\rho_{22}(T_f)$. In particular, $T_f$=100 fs and $T_f$ =100 fs+$\Delta t$ are used for the case of $\Delta t < 0$ and $\Delta t \geq 0$, respectively. To simplify the simulation, we assume that the light-matter interactions (i.e., $\vec{\mu}_{01} \cdot \vec{F}_{01}$ and $\vec{\mu}_{12} \cdot \vec{F}_{12}$), the radiative rate (i.e., $\hbar\gamma_{21}$ and $\hbar\gamma_{10}$), and the bath-induced dephasing (i.e., $\hbar\Gamma_{10}$ and $\hbar\Gamma_{21}$) of the two transitions are identical or very close. Again, according to the lifetime of the intermediate state, the values of $\hbar\gamma_{21}$ and $\hbar\gamma_{10}$ are basically in the same magnitude as 2.83 meV (1.46 ps). To reach an acceptable match of both interference fingers and enhanced TPPL beyond interference, two different frequencies ($c/(2\pi\omega_{01})$= 815 nm and $c/(2\pi\omega_{12})$= 792 nm are used for Figure 4b) are needed to couple the transition of $|0\rangle \rightarrow |1\rangle$ and $|1\rangle \rightarrow |2\rangle$, hinting two-color two-photon absorption. That is, broadband excitation is needed to prepare the coherent superposition states and to achieve super interference fringes. To verify this conclusion, we excite the same AuNP using a narrowband femtosecond laser (Figure 4c, Spirit, Spectra-Physics). The center wavelength is also 800 nm, and the repetition rate of 80 MHz, but with the pulse width of 80 fs. The resonant interference fringes excited by the narrowband laser pulse are presented in Figure 4d. We can find the enhancement factor is close to 7, much smaller than that of broadband excitation (90±14). This result indicates that only a few vibrational modes associated with the resonant interference, and thus most of TPPL intensity was originating from the non-resonant two-photon excitation. Therefore, we can expect a shorter laser pulse is crucially beneficial for the preparation of super interference fringes. Additionally, the value of $\hbar\Gamma_{20}$ can be inspired by previous works involving dephasing studies[25,38]. With these choices, the simulated result agrees fairly well with the experiment, as shown in Figure 4b. From this simulation (the corresponding parameters have been presented in Table I), we can find that the dipole moment $\vec{\mu}_{02}$ is only one order of magnitude smaller than $\vec{\mu}_{01}$ and $\vec{\mu}_{12}$, indicating the remarkable two-photon absorption cross-section of AuNP. These parameters coincide with the relevant reports. Moreover, the dephasing rate $\hbar\Gamma_{02}$ of 349 meV (11.9 fs) agrees well with the plasmon dephasing rate of similar Au nanoparticles.[25,38] These reasonable values will significantly promote our growing understanding of the quantum coherence nature of AuNP, which is a prerequisite for quantum-related applications.

To shed more light on the details of intermediate states and their critical roles on the super interference fringes, we further studied the size-dependent TPPL on Au nanospheres (AuNS) with the diameters ranging from 20 nm to 160 nm. The final interference fringes are illustrated in Figure 5a-5d, respectively. We can find that the larger the diameter, the faster the decay of interference fingers, and the smaller the enhancement factor. It can be accessible. Larger volume generally stands for more vibrational modes, which leads to a more considerable dissipation. The numerical simulations have been displayed in Figure 5e-5h; the simulations match the experiments rather well. The corresponding parameters are listed in Table I. Among these parameters, the change of $\hbar\Gamma_{01}$ ($\hbar\Gamma_{12}$) and $\hbar\Gamma_{02}$ are more significant, implying the shorter dephasing time of coherence between different states in larger AuNS. The variations of these parameters are following the increase of their vibrational modes. Note that although the volume of AuNP is slightly larger than AuNS-20, the enhancement factor of AuNP is still more remarkable than that of AuNS-20. We suggest that this is due to the much more pronounced localized surface plasmon resonance in nanobipyramids than nanospheres.

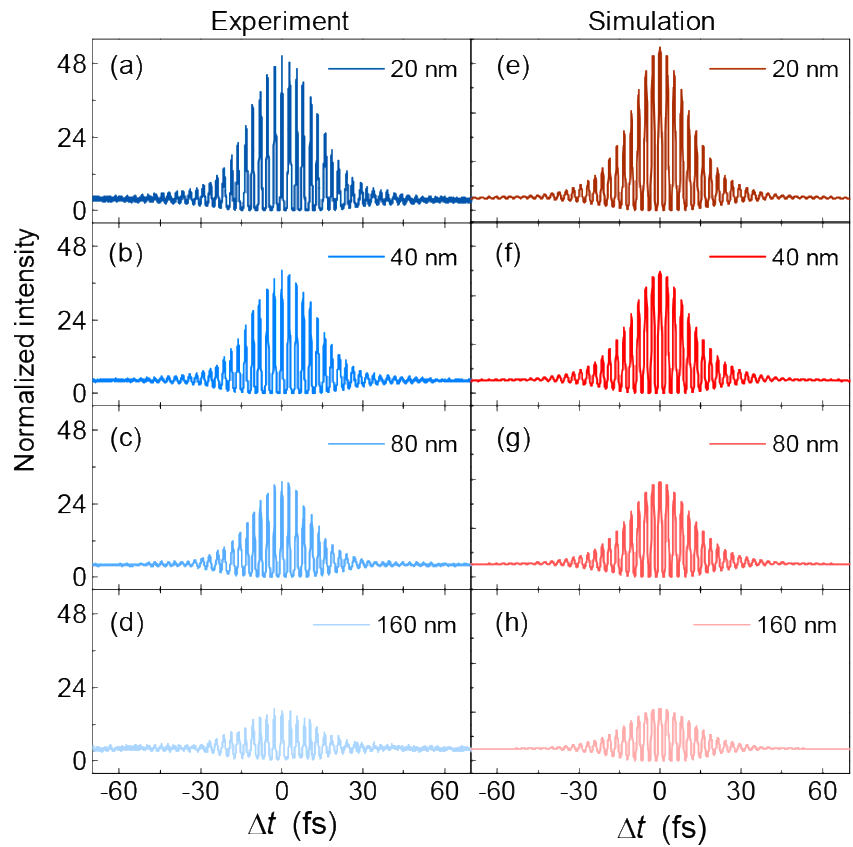

Figure 5. Size dependence of super interference fringes. (a)-(d) Experimental data for

Au nanospheres (AuNS) with a diameter of 20 nm, 40 nm, 80 nm, and 160 nm, respectively. The corresponding numerical simulations are presented in (e)-(h), respectively.

## 3. Conclusion

In summary, we have proved, based on the TPPL experiments on the time scales from sub-femtosecond to tens of picoseconds and numerical simulations with a three eigen-states model, the critical role of the intermediate state for the super interference fringes in individual AuNP. We presented that TPPL can be enhanced by almost two orders of magnitude by two-pulse excitation under resonant coherent excitation, compared to that excited by single-pulse excitations. We also demonstrated that ultrafast dynamics and power-dependent TPPL of AuNP manifested complicated but regular behaviors with the change of interpulse delays. Remarkably, the nonlinearity of TPPL ranging from 1 to 2, when varying the incident power of the first or second pulse, was determined. By considering the presence of the intermediate state and its dephasing process, all the experimental results can be well explained and simulated. We also performed the same experiments on the AuNS with different diameters. The emerging super interference fringes provide further evidence for the validity of this picture. Our finding suggests that the intermediate state dominates TPPL under two-pulse excitation and enables dramatic PL enhancement, highlighting great importance both for basic science experiments and potential practical applications in the PL imaging, lasing, and sensing well as the production of multimode coherent plasmon excitations.

Table II Parameters of the numerical simulations for different Au nanoparticles.

| Unit (meV) | $\vec{\mu}_{01} \cdot \vec{F} = \vec{\mu}_{12} \cdot \vec{F}$ | $\vec{\mu}_{01} \cdot \vec{F}_0$ | $-\hbar\Gamma_{00}$ | $\hbar\Gamma_{11}$ | $\hbar\Gamma_{22}$ | $\hbar\Gamma_{01}=\hbar\Gamma_{12}$ | $\hbar\Gamma_{02}$ |
|---|---|---|---|---|---|---|---|
| NbP | 83 | 8.3 | 15.3 | 0.1 | 15.2 | 40 | 349 |
| NP-20 | 83.6 | 8.36 | 14.7 | 0.1 | 14.6 | 42 | 349 |
| NP-40 | 86 | 8.6 | 12.3 | 0.1 | 12.2 | 50 | 349 |
| NP-80 | 87.8 | 8.78 | 10.5 | 0.1 | 10.4 | 56 | 349 |
| NP-160 | 91.7 | 9.17 | 6.6 | 0.1 | 6.5 | 69 | 349 |

**Acknowledgment**

The authors gratefully acknowledge support from the National Key Research and Development Program of China (Grant No. 2017YFA0304203), Natural Science Foundation of China (Nos. 91950109, 61875109, 61527824, and 61675119), Natural Science Foundation of Shanxi Province (No. 201901D111010(ZD)), PCSIRT (No. IRT_17R70), 111 projects (Grant No. D18001), 1331KSC, and PTIT.

**Author contributions**

C. Q., and L. X. designed and supervised the experiments. Y. L., Y. S., and S. H. carried out the optical experiments. Y. Y., and Y. L. performed the numerical simulations. G. Z., R. C., and J. H. prepared and characterized the sample. X. L., C. Q., Y. L., and Y. T. wrote the manuscript. All authors commented on the manuscript.

**Competing interests**

The authors declare no competing interests.

**Additional information**

Supplementary information is available for this manuscript.